\documentclass[twocolumn,aps,showpacs,prb,tightenlines,amsmath,amssymb,superscriptaddress,10pt,nofootinbib,nobibnotes]{revtex4-1}
\usepackage{graphicx} 
\usepackage{amssymb} 
\usepackage{dcolumn}
\usepackage{amsmath} 
\usepackage{bm} 
\usepackage{colordvi} 
\usepackage{epstopdf}
\usepackage{float} 
\usepackage{subfigure}
\usepackage{color}
\usepackage[section]{placeins}

\DeclareMathOperator\arctanh{arctanh}
\DeclareMathOperator\erf{erf}

\begin{document}

\title{On-demand electron source with tunable energy distribution}
\author{Y. Yin} 
\thanks{Author to whom correspondence should be addressed} 
\email{yin80@scu.edu.cn.} 
\affiliation{Laboratory of Mesoscopic and Low Dimensional Physics, 
  Department of Physics, Sichuan University, Chengdu, Sichuan, 610064, China} 
\date{\today}

\begin{abstract}
  We propose a scheme to manipulate the electron-hole excitation in the voltage
  pulse electron source, which can be realized by a voltage-driven Ohmic contact
  connecting to a quantum hall edge channel. It has been known that the
  electron-hole excitation can be suppressed via Lorentzian pulses, leading to
  noiseless electron current. We show that, instead of the Lorentzian pulses,
  driven via the voltage pulse
  $V(t) = 2 \frac{\hbar}{e} \sqrt{\frac{\sqrt{3}}{\pi} k_{\rm B} T_h} \arctanh(
  \frac{t - t_0}{t_0} )$ with duration $t_0$, the electron-hole excitation can
  be tuned so that the corresponding energy distribution of the emitted
  electrons follows the Fermi distribution with temperature
  $T_{\rm D} = \sqrt{ T^2_{\rm S} + T^2_{\rm h} }$, with $T_{\rm S}$ being the
  electron temperature in the Ohmic contact. Such Fermi distribution can be
  established without introducing additional energy relaxation mechanism and can
  be detected via shot noise thermometry technique, making it helpful in the
  study of thermal transport and decoherence in mesoscopic system.
\end{abstract}

\pacs{73.23.-b, 72.10.-d, 73.21.La, 85.35.Gv} 

\maketitle

\section{INTRODUCTION}
\label{sec1}

The on-demand coherent injection of single or few electrons in solid-state
circuits is an important task in electron quantum optics, which focus on the
manipulation of electrons in optics-like setups.\cite{oliver1999, henny1999,
  bertoni2000, ionicioiu2001, ji2003, bocquillon2014} The injection can be
implemented simply by a voltage-pulse-driven Ohmic contact connecting to a
quantum hall edge channel, which is usually referred as the voltage pulse
electron source.\cite{glattli2017} The Ohmic contact serves as an electron
reservoir, while the quantum hall edge channel serves as an electron
waveguide. Driven by the voltage pulse applied on the Ohmic contact, electrons
incoming from the reservoir can be injected on the Fermi sea of the edge
channel, leading to the single-electron quasi-particle excitation propagating
along the waveguide. However, additional electron-hole excitation can usually be
created during the injection,\cite{vanevic2016} inducing charge current noise.

As far as the charge transport is concerned, it is desired to suppress the
electron-hole excitation. In a series of seminal works, Levitov {\em et al.}
have proposed that, driven by Lorentzian pulses with integer Faraday flux,
integer electrons can be injected, while the accompanied electron-hole
excitation can be suppressed, leading to a noiseless current flow.\cite{lee1993,
  lee1995, keeling2006} This has been realized and extensively studied in the
experiments reported in the group of D.C. Glattli.\cite{dubois2013, jullien2014,
  glattli2017} Later, Gabelli {\em et al.} further show that a similar
suppression can be realized by using bi-harmonic voltage
pulses.\cite{gabelli2017} Besides, the existence of the electron-hole excitation
can also be helpful in certain situation. Moskalets have demonstrated that,
driven by the Lorentzian pulse with a half-integer flux, a zero-energy
excitation with half-integer charge can be created in the driven Fermi sea at
zero temperature, which cannot exist without the presence of electron-hole
excitation.\cite{moskalets2016} These works demonstrate the possibility of
manipulating electron-hole excitation via engineering the temporal profile of
the voltage pulse.

In contrast, if the thermal transport is concerned,\cite{schwab2000,
  chiatti2006, chen2009, granger2009, altimiras2010, le2010, venkatachalam2012,
  Jezouin2013} the electron-hole excitation are favorable, since they carry a
finite amount of energy, while do not affect the average charge transport. In
fact, a fully thermalized state at finite temperature can be regarded as the
mixed state of certain electron-hole excitations, where the energy distribution
follows the Fermi distribution, while the quantum coherence totally vanishes. If
the electron-hole excitation can be manipulated via engineering the voltage
pulse, it is then nature to ask: is it possible to tune the electron-hole
excitation so that the corresponding state has exactly the same energy
distribution of the fully thermalized state, while the quantum coherence is
still preserved?  Such a state will be helpful in the study of thermal transport
and quantum decoherence processes in mesoscopic systems.

In this paper, we propose a scheme to create such a state by using the voltage
pulse electron source. We find that, instead of Lorentzian pulses, driving via
the voltage pulse with duration $t_0$, i.e., $t \in [0, t_0]$, which has the
temporal profile
\begin{eqnarray}
  V(t) & = & 2 \frac{\hbar}{e} \sqrt{\frac{\sqrt{3}}{\pi} k_{\rm B} T_h}
             \arctanh( \frac{t - t_0}{t_0} ),
  \label{s1:eq1}
\end{eqnarray}
with $k_{\rm B}$ being the Boltzmann constant, electrons incoming from the
reservoir at temperature $T_{\rm S}$ and chemical potential $\mu$ can be excited
so that the corresponding outgoing electrons can have the time-averaged energy
distribution as
\begin{eqnarray}
  f_{\rm D}(\omega) & = & \frac{1}{1+\exp[(\hbar \omega -\mu)/(k_{\rm B} T_{\rm D})]},
                             \label{s1:eq2}
\end{eqnarray}
which is exactly a Fermi distribution at temperature
\begin{eqnarray}
  T_{\rm D} & = & \sqrt{ T^2_{\rm S} + T^2_{\rm h} }.
                  \label{s1:eq3}
\end{eqnarray}

Note that such Fermi distribution is obtained solely by coherent excitation via
voltage pulses and no additional energy relaxation mechanism is needed. Hence
the quantum coherence is still preserved for the state. Experimentally, it is
possible to detect the energy distribution of the state via quantum-dot-based
energy filter.\cite{altimiras2010, le2010} We further show that such state can
also be detected {\em in situ} at the voltage pulse electron source via shot
noise thermometry technique,\cite{spietz2003} making it convenient in further
experimentally studies.

The paper is organized as follows. In Sec.~\ref{sec2}, we present the model of
the voltage pulse electron source and introduce the time-averaged energy
distribution for the electrons. In Sec.~\ref{sec3}, we demonstrate how to
manipulate the temporal profile of the pulse so that the time-averaged energy
distribution of the emitted electrons follows the Fermi distribution we
desired. In Sec.~\ref{sec4}, we discuss the detection of the state via shot
noise thermometry technique. We summarized in Sec.~\ref{sec5}.

\section{Model and Formalism}
\label{sec2}

\begin{figure}
  \centering
  \includegraphics[width=7.5cm]{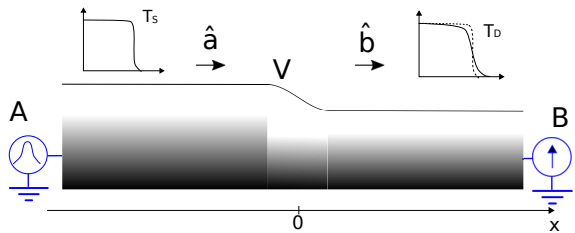}
  \caption{(Color online) Schematic of the voltage pulse source, composing of an
    one-dimensional quantum wire connecting two reservoirs A and B. Electrons
    are injected from the reservoir A by the time-dependent bias voltage
    $V$. The current and their fluctuation are detected in the reservoir B. The
    incoming(outgoing) wave-packet from(towards) the reservoir A(B) is
    represented by the operator $\hat{a}$($\hat{b}$). The energy distribution of
    the incoming (at temperature $T_{\rm S}$) and outgoing (at temperature
    $T_{\rm D}$) electron wave-packet are also illustrated. The 1D coordinates
    system is shown in the bottom of the figure.}
  \label{fig1}
\end{figure}

The voltage pulse electron source can be modeled as an one-dimensional quantum
wire connecting two reservoir A and B,\cite{keeling2006} as illustrated in
Fig.~\ref{fig1}. The system is biased with a time-dependent voltage pulse $V(t)$
with duration $t_0$. The voltage drop is assumed to occur across a short
interval at the center of the wire. If the voltage drop is spatially
slow-varying on the scale of the Fermi wavelength $1/k_{\rm F}$, the electron in
such system can be well-approximated as dispersionless Fermi systems with the
corresponding single-particle Hamiltonian
\begin{eqnarray}
  H & = & - i \hbar ( \pm v_{\rm F} ) \partial_x + e V(t) \theta(-x),
          \label{s2:eq1}
\end{eqnarray}
with $e$ being the charge of electron and $v_{\rm F}$ being the Fermi
velocity. The sign $\pm$ corresponds to left- and right- going electrons,
respectively. Without loss of generality, we focus on the right-going electron.

The field operator of the electron $\hat{\psi}(x,t)$ can be expressed as
\begin{eqnarray}
  \hat{\psi}(x,t) & = & \left\{
                  \begin{tabular}{cc}
                    $\hat{a}(t-x/v_{\rm F}) e^{-i \phi(t)}$ &,~$x<0$  \\
                    $\hat{b}(t-x/v_{\rm F})$ &,~$x>0$  \\
                  \end{tabular}
  \right.,
  \label{s2:eq2}
\end{eqnarray}
with $\phi(t) = \frac{e}{\hbar} \int^t d\tau V(\tau)$ describing the effect of
the voltage pulse $V(t)$. The Fermi operator $\hat{a}(t)$ and $\hat{b}(t)$
corresponds to the incoming and outgoing electron modes, respectively, which is
also indicated in Fig.~\ref{fig1}. They are related via the forward scattering
phase\cite{blanter2000, keeling2006, ivanov1997, keeling2008}
\begin{eqnarray}
  \hat{b}(t) & = & \hat{a}(t) e^{-i \phi(t)},
  \label{s2:eq3}
\end{eqnarray}
which can be obtained by requiring the field operator $\hat{\psi}(x,t)$ is
continuity at the boundary $x=0$.

Now we turn to discuss the energy distribution function of the incoming and
outgoing electrons. Following previous works, the non-interacting electrons can
be characterized by their one-body density matrix, which has the form of the
first-order Glauber correlation function in the time-domain.\cite{grenier2011,
  haack2011, haack2013} The correlation function for the incoming and outgoing
electrons can be constructed as
\begin{eqnarray}
  G_a(t, t') & = & \left< \hat{a}^{\dagger}(t) \hat{a}(t') \right>, \nonumber\\
  G_b(t, t') & = & \left< \hat{b}^{\dagger}(t) \hat{b}(t') \right>, 
                   \label{s2:eq4}
\end{eqnarray}
respectively, where $\left< ... \right>$ represents the thermal expectation over
the reservoir degree of freedom.

For incoming electrons, which can be modelled as the stationary wave-packet
emitted from the reservoir A at thermal equilibrium, the correlation function
satisfies translation invariant in the time-domain, and hence only depends on
the time difference $\tau=t-t'$. The energy distribution function can be
obtained through Fourier transformation, which has the form
\begin{eqnarray}
  f_a(\omega) & = & \int d\tau e^{i \omega \tau} G_a(t, t-\tau).
                               \label{s2:eq5}
\end{eqnarray}
For reservoir A at temperature $T_{\rm S}$ and chemical potential $\mu$, the
distribution $f_a(\omega)$ follows the Fermi distribution
\begin{eqnarray}
  f_{\rm S}(\omega) & = & \frac{1}{1+\exp[(\hbar \omega -\mu)/(k_{\rm B} T_{\rm S})]}.
                               \label{s2:eq6}
\end{eqnarray}

In contrast, driven by the time-dependent potential $V(t)$, the outgoing
electrons are described by the non-stationary wave-packet and the translation
invariant of the correlation function is broken. In this situation, one can
introduce the time-averaged energy distribution function over the time interval
$t_0$, which can be written as\cite{lee1993, lee1995, kovrizhin2011,
  kovrizhin2012, moskalets2014, moskalets2016}
\begin{eqnarray}
  f_{\rm V}(\omega) & = & \frac{1}{t_0} \int^{t_0}_0 \int^{t_0}_0 dtdt' e^{i \omega (t-t')} G_b(t,t'),
                               \label{s2:eq7}
\end{eqnarray}
which can be measured experimentally by using quantum dot as an adjustable
energy filter.\cite{altimiras2010, kovrizhin2011, kovrizhin2012} By using the
scattering phase given in Eq.~\eqref{s2:eq3}, the time-averaged distribution
function of the outgoing electrons can be related to the incoming ones via
\begin{eqnarray}
  f_{\rm V}(\omega) & = & \int \frac{d \omega'}{2 \pi} f_{\rm S}(\omega') \Pi_{\rm V} (\omega - \omega'), \nonumber\\
  \Pi_{\rm V} (\omega) & = & | \frac{1}{\sqrt{t_0}} \int^{t_0}_0 dt e^{i
                       \omega t} e^{-i \phi(t)} |^2.
                       \label{s2:eq8}
\end{eqnarray}

\section{Tuning distribution via voltage pulse}
\label{sec3}

\begin{figure}
  \centering
  \includegraphics[width=6.5cm]{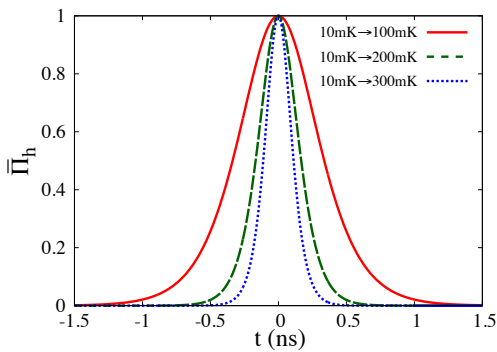}
  \caption{(Color online) Schematic of the function $\bar{\Pi}_h(t)$ for
    different temperatures. The red solid, green dashed and blue dotter curves
    represent the heating up from $T_{\rm S}=10$mK to $T_{\rm D}=100$mK, $200$mK
    and $300$mK, respectively.}
  \label{fig2}
\end{figure}

If the distribution $f_{\rm V}(\omega)$ given in Eq.~\eqref{s2:eq8} follows the
Fermi distribution $f_{\rm D}(\omega)$ given in Eq.~\eqref{s1:eq2}, then the
outgoing electron state will have the same energy distribution as a fully
thermalized state at the temperature $T_{\rm D}$. To do so, one requires that
the integral kernel $\Pi_{\rm V}(\omega)$ in Eqs.~\eqref{s2:eq8} equals to(see
Appendix~\ref{app1} for the derivation)
\begin{eqnarray}
  \Pi_h(\omega) & = & \int dt e^{i \omega t} \bar{\Pi}_h(t),
                      \label{s3:eq1}
\end{eqnarray}
with
\begin{eqnarray}
  \bar{\Pi}_h(t) & = & \frac{T_{\rm D}}{T_{\rm S}} \frac{\sinh(\frac{\pi k_{\rm B} T_{\rm S}
                       }{\hbar} t)}{\sinh(\frac{\pi k_{\rm B} T_{\rm D}}{\hbar
                       } t)}.
                       \label{s3:eq1.1}
\end{eqnarray}
The typical profiles of $\bar{\Pi}_h(t)$ at sub-kelvin temperatures are
illustrated in Fig.~\ref{fig2}.

To achieve such requirement, one need to find a proper $V(t)$ so that the power
spectral density of $e^{-i \phi(t)}$ follows $\Pi_h(\omega)$, i.e.,
\begin{eqnarray}
  \Pi_h(\omega) & = \Pi_{\rm V}(\omega) = & | \frac{1}{\sqrt{t_0}} \int^{t_0}_0 dt e^{i \omega t} e^{-i \phi(t)} |^2,
  \label{s3:eq2}
\end{eqnarray}
with $\phi(t)$ given below Eq.~\eqref{s2:eq3}. It should be noted that since
$\Pi_{\rm V}(\omega)$ is non-negative according to Eq.~\eqref{s2:eq8},
equation~\eqref{s3:eq2} cannot be satisfied for $T_{\rm D} < T_{\rm S}$, since
$\Pi_h(\omega)$ can be negative in this case.

Finding $V(t)$ satisfying the requirement Eq.~\eqref{s3:eq2} is equivalent to
the problem of phase control of pulse shaping, which has been extensively
studied in the field of ultrafast optics.\cite{weiner2011} The basic idea is to
attack the problem by working out the Fourier transformation in
Eq.~\eqref{s3:eq2} within the stationary phase approximation. Here we only
outline the procedure, leaving the technical details to the Appendix~\ref{app2}.

Within the stationary phase approximation, Eq.~\eqref{s3:eq2} can be satisfied
by requiring the voltage pulse $V(t)$ follows the relation
\begin{eqnarray}
  \frac{t}{t_0} & = & \int^{\frac{e}{\hbar}V(t)}_{-\infty} \frac{d \omega}{2\pi} \Pi_h(\omega),
  \label{s3:eq3}
\end{eqnarray}
with $t \in [0, t_0]$. An analytical solution of Eq.~\eqref{s3:eq3} can be
obtained by approximating $\Pi_h(\omega)$ via the Gaussian ansatz[see
Appendix~\ref{app1} for details]
\begin{eqnarray}
  \Pi_h(\omega) & \approx & \frac{\sqrt{2\pi}}{\sigma_h} e^{-\frac{1}{2} ( \frac{\omega}{\sigma_h} )^2},
  \label{s3:eq4}
\end{eqnarray}  
where $\sigma_h = \pi k_{\rm B} T_h/\sqrt{3}$ with $T_h = \sqrt{T^2_{\rm D} - T^2_{\rm
    S}}$.

Substituting Eq.~\eqref{s3:eq4} into Eq.~\eqref{s3:eq3}, one has
\begin{eqnarray}
  V(t) & = & \sqrt{2} \sigma_h \erf^{-1}(\frac{2t}{t_0}).
  \label{s3:eq5}
\end{eqnarray}  
By further applying the approximation
$\erf^{-1}(x) \approx \frac{\sqrt{6}}{\pi} \arctanh(x)$, the above equation
reduced to
\begin{eqnarray}
  V(t) & = & 2 \frac{\hbar}{e} \sqrt{\frac{\sqrt{3}}{\pi} k_{\rm B} T_h}
             \arctanh( \frac{t - t_0}{t_0} ),
  \label{s3:eq6}
\end{eqnarray}
which is just Eq.~\eqref{s1:eq1} given in Sec.~\ref{sec1}.

Note that there are some freedom in the choice of the pulse duration $t_0$. In
principle, $t_0$ should be large enough so that the stationary phase
approximation holds. In the realistic calculation, we find that it is adequate
to choose $t_0$ to be two times larger than the Full width at half maximum
(FWHM) of the profile $\bar{\Pi}_h(t)$. According to Fig.~\ref{fig2}, the
typical value of $t_0$ is of the order of $1$~ns at sub-kelvin temperatures,
corresponding to the frequency of the order of $1$~GHz. One may also worry about
the divergence of the function $\arctanh(\tau)$ at the boundary $\tau=\pm 1$ in
Eq.~\eqref{s3:eq6}. This can be fixed by set a limit of the amplitude of $V(t)$,
which has minor influence if the overall profile of $V(t)$ does not change too
much.

\begin{figure}
  \centering
  \includegraphics[width=7.5cm]{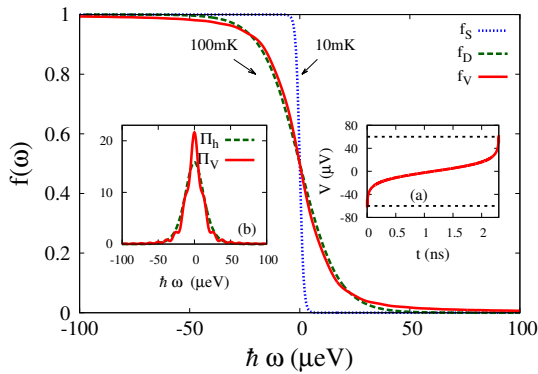}
  \caption{(Color online) The energy distribution of the incoming and outgoing
    electrons (main panel). The blue dotted curves represents the distribution
    of the incoming electron $f_{\rm S}$, which is the Fermi distribution at the
    temperature $T_{\rm S}=10$~mK. The distribution of the outgoing electrons
    $f_{\rm V}$ are shown with red solid curves. The green dashed curve
    represent the Fermi distribution at the temperature $T_{\rm D}=100$~mK. The
    voltage pulse $V(t)$ with duration $t_0=2.3$~ns is shown in inset (a), the
    amplitude of $V(t)$ is limited to $60$~$\mu$V as indicated by the black
    dotted lines. The integral kernel $\Pi_{\rm V}(\omega)$ corresponding to
    $V(t)$(red solid curve) is compared to the integral kernel $\Pi_h(\omega)$
    from Eq.~\eqref{s3:eq1}(green dashed curve) in the inset (b).}
  \label{fig3}
\end{figure}

Despite the approximation we have used in above derivation, the time-averaged
energy distribution of the outgoing electron produced by the voltage pulse
$V(t)$ can follow the desired Fermi distribution quite well, which we
demonstrate in Fig.~\ref{fig3}. In the main panel of the figure, the blue dotted
curves represents the energy distribution $f_{\rm S}(\omega)$ of the incoming
electrons from reservoir A, which is the Fermi distribution at the temperature
$T_{\rm S}=10$~mK and chemical potential $\mu=0$. The red solid curve represents
the time-averaged distribution $f_{\rm V}(\omega)$ of the outgoing electrons,
which is calculated by numerical integrating Eqs.~\eqref{s2:eq8} with the
voltage pulse $V(t)$[Eq.~\eqref{s3:eq6}] of the parameter $T_h=99.5$~mK. One can
see that it follows the Fermi distribution at the temperature
$T_{\rm D}=100$~mK(green dashed curve), which agrees with the analytical
expression given in Eq.~\eqref{s3:eq6}.

The temporal profile of the applied voltage pulse $V(t)$ is shown in the inset
(a) of Fig.~\ref{fig3}. The pulse duration $t_0$ is chosen to be $2.3$~ns, which
is just equal to two times of the FWHM of the corresponding $\bar{\Pi}_h(t)$[red
solid curve in Fig.~\ref{fig2}]. The amplitude limit of $V(t)$ is set to
$60$~$\mu$V, which has little impact on the overall profile of the pulse. In the
inset (b), we also compare the integral kernel $\Pi_{\rm V}(\omega)$[calculated
from $V(t)$ via Eqs.~\eqref{s2:eq8}] to the integral kernel $\Pi_h(\omega)$
given in Eq.~\eqref{s3:eq1}. One can see that, despite some small ripples, the
overall profile of the two kernel agrees with each other, which justifies the
approximation we have used in the derivation.

\begin{figure}
  \centering
  \includegraphics[width=7.5cm]{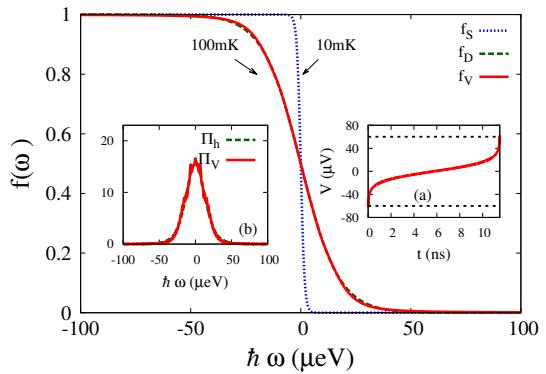}
  \caption{(Color online) The same as Fig.~\ref{fig3}, but with the pulse
    duration $t_0=11.5$~ns.}
  \label{fig4}
\end{figure}

By further increasing the pulse duration $t_0$, the ripples in the integral
kernel can be suppressed, as can be seen from Fig.~\ref{fig4}. Comparing to
Fig.~\ref{fig3}, we have increase $t_0$ to $11.5$~ns, while keeping other
parameter fixed. The suppression of the ripples in the integral kernel can be
clearly in the inset (b) of Fig.~\ref{fig4}. As a consequence, the energy
distribution of the outgoing electrons(red solid curve) agrees quite well with
the Fermi distribution with $T_{\rm D}=100$~mK(green dashed curve), so that they
are almost indistinguishable from each other in the main panel of the figure.

Hence, one can see that, applying the voltage pulse $V(t)$ with the temporal
profile given in Eq.~\eqref{s3:eq6}, the energy distribution of the outgoing
electrons can be tuned to follow the Fermi distribution from Eq.~\eqref{s1:eq2}
quite well. According to Eq.~\eqref{s3:eq6}, to achieve such tuning for higher
temperature $T_{\rm D}$, voltage pulse with larger amplitude is required. In
this case, the amplitude limit of the pulse can have more pronounced impact on
the result.

\begin{figure}
  \centering
  \includegraphics[width=7.5cm]{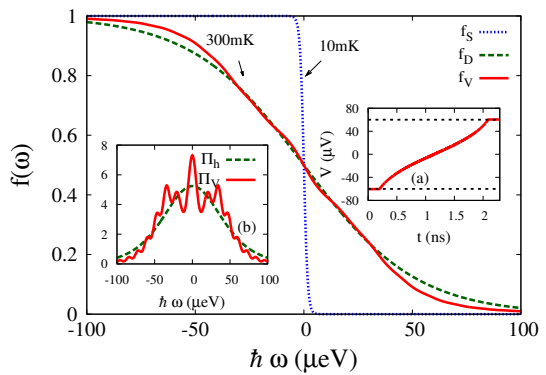}
  \caption{(Color online) The same as Fig.~\ref{fig3}, but with the temperature
    $T_{\rm D}=300$~mK and the corresponding $T_h=299.8$~mK.}
  \label{fig5}
\end{figure}

Such impact is demonstrated in Fig.~\ref{fig5}. All the parameters are chosen
the same as Fig.~\ref{fig3}, except the temperature $T_{\rm D}$, which is
increased to $300$~mK in this case. From the inset (a), one can see that the
pulse profile is modified due to the amplitude limit of the pulse, leading to
small platforms close to the edges. As a consequence, these platforms induces
ripples in the corresponding integral kernel, which is shown in the inset
(b). Such ripples can also be identified in the energy distribution of the
outgoing electron, which is shown as red solid curve in the main panel of the
figure.

\begin{figure}
  \centering
  \includegraphics[width=7.5cm]{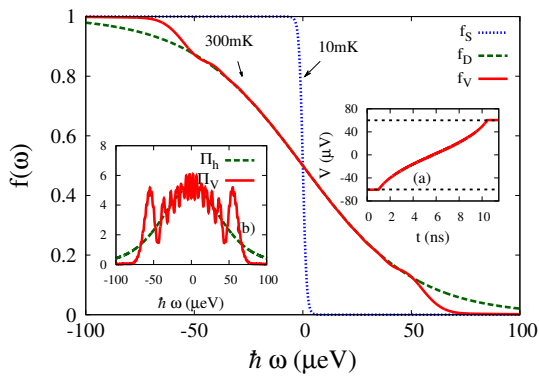}
  \caption{(Color online) The same as Fig.~\ref{fig5}, but with the pulse
    duration $t_0=11.5$~ns.}
  \label{fig6}
\end{figure}

Note that in the case where the profile of the pulse has been changed due to the
amplitude limitation, the induced ripples cannot be suppressed by increasing the
pulse duration $t_0$, which we demonstrate in Fig.~\ref{fig6}. In this case, all
the parameters are chosen the same as Fig.~\ref{fig5}, except the pulse duration
$t_0$, which is increased to $11.5$~ns. One can still identify large ripples in
the integral kernel[inset (b)]. The impact of these ripples can lead to
pronounced distortion of the energy distribution of outgoing electrons far from
the Fermi level, which can be seen in the main panel of Fig.~\ref{fig6}.

It should be noted that in the realistic situation, the voltage pulse can also
lead to Joule heating.\cite{kumar1996} To make the manipulation observable, the
amplitude of the voltage pulse should be kept low enough so that the Joule
heating is not significant. It has been reported that in typical quantum point
contacts, voltage pulses with the root mean square amplitude $60$~$\mu$V can
heat up the electrons to about $50$~mK,\cite{dubois2013} which is smaller then
the effect we shown here. So we expect that such manipulation can be
experimentally detected in a realistic system.

\section{Shot noise thermometer}
\label{sec4}

Experimentally, the time-averaged energy distribution of the outgoing electrons
can be detected via an adjustable energy filter, which can be realized by using
a quantum dot fabricated at a certain distance from the voltage
source.\cite{altimiras2010, kovrizhin2011} However, as the profile of the energy
distribution can be changed during the propagation due to various energy
relaxation mechanisms, such as the coupling to electrons on counter-propagating
edge channels,\cite{le2010, kovrizhin2012, venkatachalam2012, prokudina2014} it
is more favorable to detect the distribution {\em in situ} at the voltage pulse
source. This can be done by using the shot noise thermometry
technique,\cite{spietz2003} which we will discuss in this section.

\begin{figure}
  \centering
  \includegraphics[width=7.5cm]{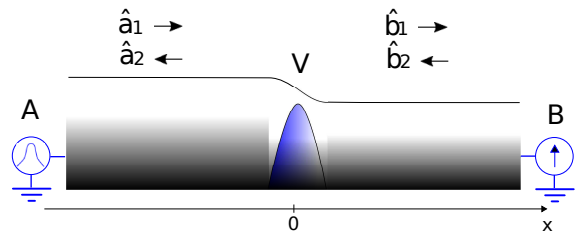}
  \caption{(Color online) Schematic of the voltage pulse source with a static
    potential barrier, which mixes the left-going (represented by $\hat{a}_1$
    and $\hat{b}_1$) and right-going (represented by $\hat{a}_2$ and
    $\hat{b}_2$) electrons.}
  \label{fig7}
\end{figure}

In the standard shot noise thermometry technique, the electron temperature is
extracted from the DC-bias-voltage dependence of the current noise across a
tunneling barrier.\cite{spietz2003} This can be implemented in the voltage pulse
electron source by introducing a static potential in the short interval at the
center of the wire, which is illustrated in Fig.~\ref{fig7}. Note that the
static potential is assumed to be rapidly-varying comparing to the Fermi
wavelength $1/k_{\rm F}$, which can lead to backscattering between left- and
right-going electrons in the system.

Following the scattering formalism, the scattering between the left- and
right-going electrons can be described by the time-dependent scattering
matrix\cite{blanter2000, ivanov1997}
\begin{eqnarray}
  && \left(
     \begin{tabular}{c}
       $\hat{b}_1(t)$\\
       $\hat{a}_2(t)$\\
     \end{tabular}
  \right) = S(t) \left(
  \begin{tabular}{c}
    $\hat{a}_1(t)$\\
    $\hat{b}_2(t)$\\
  \end{tabular}
  \right), \nonumber\\
  && S(t) = \left(
     \begin{tabular}{cc}
       $\sqrt{D_0} e^{-i \phi(t)}$& $-i \sqrt{1-D_0}$ \\
       $-i \sqrt{1-D_0}$& $\sqrt{D_0} e^{i \phi(t)}$ \\            
     \end{tabular}
  \right), 
  \label{s4:eq1}
\end{eqnarray}
with $D_0$ being the time-independent transmission coefficient due to the static
potential. The Fermi operator $\hat{a}_{1(2)}$ and $\hat{b}_{2(1)}$ represent
the incoming(outgoing) electrons modes from(towards) the reservoir A and B,
respectively. The phase factor $\phi(t)$ is given below Eq.~\eqref{s2:eq3}. For
simplicity, we assume both the reservoir A and B are kept at the same
temperature $T_{\rm S}$ and chemical potential $\mu$.

Consider the time-averaged current noise detected at the reservoir B, it can be
given as
\begin{eqnarray}
  S_{\rm B} & = \frac{1}{t_0}& \int^{t_0}_0 dt \int^{t_0}_0 dt' \Big[ \left< \hat{j}_{\rm
                  B}(t) \hat{j}_{\rm B}(t') \right> \nonumber\\
  && \hspace{1.5cm}\mbox{}- \left< \hat{j}_{\rm B}(t)
                  \right> \left< \hat{j}_{\rm B}(t') \right> \Big],
               \label{s4:eq11}
\end{eqnarray}
where the current operator $\hat{j}_{\rm B}(t)$ has the form
\begin{eqnarray}
  \hat{j}_{\rm B}(t) & = & \hat{b}^{\dagger}_1(t) \hat{b}_1(t) - \hat{b}^{\dagger}_2(t) \hat{b}_2(t).
               \label{s4:eq12}
\end{eqnarray}
By using the time-dependent scattering matrix Eq.~\eqref{s4:eq1}, the current
noise can be written as
\begin{eqnarray}
  S_B & = & S_e + S_n, 
  \label{s4:eq2}
\end{eqnarray}
with $S_e$ and $S_n$ being the equilibrium and non-equilibrium contribution,
respectively. The equilibrium term
\begin{eqnarray}
  S_e & = & \frac{e^2}{2\pi \hbar} D^2_0 \int \frac{d\omega}{2\pi} \Big\{ \big[ 1 - f_{\rm S}(\Omega)
            \big] f_{\rm S}(\omega) \nonumber\\
      && \hspace{1.5cm}\mbox{}+ f_{\rm S}(\omega) \big[ 1 - f_{\rm S}(\omega) \big] \Big\},
         \label{s4:eq3}
\end{eqnarray}
is just the Nyquist-Johnson noise which is independent on the applied DC bias
voltage. The non-equilibrium term, i.e., the shot noise term, can be written as
\begin{eqnarray}
  S_n & = & \frac{e^2}{2\pi \hbar} D_0(1-D_0) \int \frac{d\omega}{2\pi} \Big\{ \big[ 1 - f_{\rm S}(\omega) \big]
            f_{\rm V}(\omega) \nonumber\\
      && \hspace{1.5cm}\mbox{}+ f_{\rm V}(\omega) \big[ 1 - f_{\rm S}(\omega)
         \big] \Big\},
         \label{s4:eq4}
\end{eqnarray}
with $f_{\rm V}(\omega)$ given in Eqs.~\eqref{s2:eq8}.

One can see immediately that, if $f_{\rm V}(\omega)$ is tuned to the Fermi
distribution $f_{\rm D}(\omega)$ given in Eq.~\eqref{s1:eq2}, then
Eq.~\eqref{s4:eq4} should be equal to
\begin{eqnarray}
  \bar{S}_n & = & \frac{e^2}{2\pi \hbar} D_0(1-D_0) \int \frac{d\omega}{2\pi} \Big\{ \big[ 1 - f_{\rm S}(\omega) \big]
                  f_{\rm D}(\omega) \nonumber\\
            && \hspace{1.5cm}\mbox{}+ f_{\rm S}(\omega) \big[ 1 - f_{\rm D}(\omega)
               \big] \Big\},
         \label{s4:eq6}
\end{eqnarray}
which is just the shot noise between two reservoirs with different temperature
$T_{\rm S}$ and $T_{\rm D}$. Hence, to check if the manipulation of the energy
distribution is properly achieved, one just compare the DC bias dependence $S_n$
from Eq.~\eqref{s4:eq4} to the predicted one $\bar{S}_n$ from
Eq.~\eqref{s4:eq6}.

\begin{figure}
  \centering
  \includegraphics[width=7.5cm]{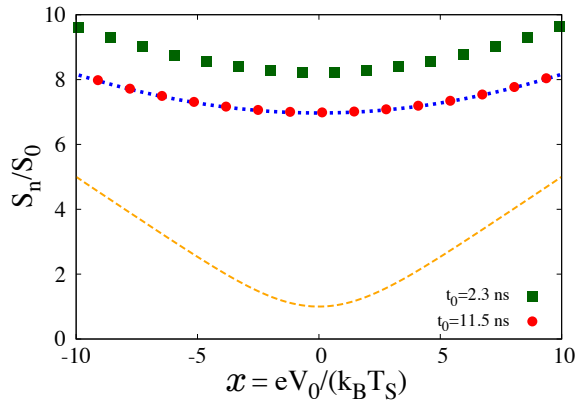}
  \caption{(Color online) Normalized shot noise $S_n/S_0$ as a function of
    normalized DC bias voltage $x = eV_0/(k_{\rm B} T_{\rm S})$, with
    $S_0= e^2 D_0 (1-D_0) k_{\rm B} T_{\rm S}/(2 \pi^2 \hbar)$ being the
    zero-bias noise. The temperature $T_{\rm S}$ of the reservoirs A and B is
    chosen to be $T_{\rm S}=10$~mK. The green squares and red dots are shot
    noise with voltage pulse duration $t_0=2.3$~ns and $11.5$~ns, corresponding
    to the case in Fig.~\ref{fig3} and Fig.~\ref{fig4}, respectively. The blue
    dotted curve represents the normalized shot noise between two reservoirs at
    temperature $T_{\rm S}$ and $T_{\rm D}$. The orange dashed curve represents
    the universal function $x/\tanh(x/2)$.}
  \label{fig8}
\end{figure}

As an example, we perform such comparison for the case with $T_{\rm S}=10$~mK
and $T_{\rm D}=100$~mK in Fig.~\ref{fig8}. The shot noise is normalized to the
zero-bias noise $S_0= e^2 D_0 (1-D_0) k_{\rm B} T_{\rm S}/(2 \pi^2 \hbar)$,
which is plotted as a function of the normalized voltage
$x = eV_0/(k_{\rm B} T_{\rm S})$. The predicted $\bar{S}_n$ from
Eq.~\eqref{s4:eq6} is plotted as blue dotted curve. The green squares represent
the shot noise obtained by applying the voltage pulse $V(t)$ with $t_0=2.3$~ns,
while the red dots correspond to the case with $t_0=11.5$~ns. The corresponding
energy distribution of the outgoing electrons can be found in
Fig.~\ref{fig3}[$t_0=2.3$~ns] and Fig.~\ref{fig4}[$t_0=11.5$~ns],
respectively. One can see that, by increasing $t_0$, the profile of the
distribution of outgoing electrons approaches the Fermi distribution
$f_{\rm D}(\omega)$[Fig.~\ref{fig3} and Fig.~\ref{fig4}] , while the shot noise
approaches the predicted shot noise $\bar{S}_n$, indicating that the shot noise
can be used as a hallmark of the profile of the energy distribution of the
outgoing electrons. Note that without the voltage pulse $V(t)$, the normalized
shot noise follows the well-known universal function $x/\tanh(x/2)$, which is
shown as orange dashed curves in the figure.

\section{SUMMARY}
\label{sec5}

In summary, we have propose a scheme to manipulate the electron-hole excitation
in the voltage pulse electron source. By using stationary phase approximation,
we derive a simple analytical expression of the voltage pulse $V(t)$
[Eq.~\eqref{s1:eq1}], which can tuned the electron-hole excitation so that the
energy distribution of the emitted electrons from the electron source follows a
desired Fermi distribution with higher temperature. Such distribution can be
established without introducing additional energy relaxation mechanism. We also
show that such distribution can be detected {\em in situ} at the voltage pulse
electron source via shot noise thermometry technique, making it helpful in the
study of thermal transport and decoherence in mesoscopic system.

\begin{acknowledgments}
  The author would like to thank Professor J. Gao for bringing the problem to
  the author’s attention. This work was supported by Key Program of National
  Natural Science Foundation of China under Grant No. 11234009, the National Key
  Basic Research Program of China under Grant No. 2016YFF0200403,  and Young
  Scientists Fund of National Natural Science Foundation of China under Grant
  No. 11504248.
\end{acknowledgments}

\appendix

\section{Derivation of Eq.~\eqref{s3:eq1},~\eqref{s3:eq1.1}  and~\eqref{s3:eq4}}
\label{app1}

In this appendix, we derive the integral kernel $\Pi_h(\omega)$ given in
Eqs.~\eqref{s3:eq1} and~\eqref{s3:eq1.1}. It satisfies the condition 
\begin{eqnarray}
  f_{\rm D}(\omega) & = & \int \frac{d \omega'}{2 \pi} f_{\rm S}(\omega')
                          \Pi_{\rm h} (\omega - \omega'),
                          \label{a1:eq1}
\end{eqnarray}
with $f_{\rm D}(\omega)$ and $f_{\rm S}(\omega)$ being the Fermi distribution
given in Eq.~\eqref{s1:eq2} and Eq.~\eqref{s2:eq6}, respectively.

This equation can be solved via Fourier transformation. Taking the derivative
with respect to $\omega$ on both side of Eq.~\eqref{a1:eq1}, one has
\begin{eqnarray}
  \frac{\beta_{\rm D}}{4} \frac{1}{\cosh^2[\frac{\beta_{\rm D}(\mu -
  \hbar\omega)}{2}]} & = & \int \frac{d \omega'}{2 \pi} \Pi_{\rm h} (\omega -
                           \omega') \nonumber\\
                     && \mbox{} \times \frac{\beta_{\rm S}}{4} \frac{1}{\cosh^2[\frac{\beta_{\rm S}(\mu - \hbar\omega)}{2}]},
                          \label{a1:eq2}
\end{eqnarray}
with $\beta_{\rm S(D)} = 1/[ k_{\rm B} T_{\rm S(D)} ]$.

Performing the Fourier transformation on both side of Eq.~\eqref{a1:eq2} and
using the identity 
\begin{eqnarray}
  \int dx \frac{e^{i x y}}{\cosh^2(x)} & = & \frac{\pi y}{\sinh(\frac{\pi}{2}y)},
  \label{a1:eq3}
\end{eqnarray}
one obtains 
\begin{eqnarray}
   \frac{t}{ 2\beta_{\rm D} \sinh(\frac{\pi}{\beta_{\rm D}} t)} & = & \frac{t}{
                                                                      2\beta_{\rm
                                                                      S}
                                                                      \sinh(\frac{\pi}{\beta_{\rm
                                                                      S}} t)} \bar{\Pi}_h(t),
  \label{a1:eq4}
\end{eqnarray}
with
\begin{eqnarray}
  \Pi_h(\omega) & = & \int dt e^{i \omega t} \bar{\Pi}_h(t),
                      \label{a1:eq5}
\end{eqnarray}
which gives $\Pi_h(\omega)$ given in Eq.~\eqref{s3:eq1} and~\eqref{s3:eq1.1}.

One can also apply a Gaussian approximation for the derivative of the
distribution function. In this case, one assume
\begin{eqnarray}
  f_{\rm S(D)}(E) & = &  1 - \int^E_0 dE' g_{\rm S(D)}(E'),
  \label{a1:eq6}
\end{eqnarray}
with
\begin{eqnarray}
  g_{\rm S(D)}(E) & = & \frac{1}{\sqrt{2\pi}\sigma_{\rm S(D)}} e^{-\frac{1}{2}
             (\frac{E-\mu}{\sigma_{\rm S(D)}})^2 },
  \label{a1:eq7}
\end{eqnarray}
where the width $\sigma_{\rm S(D)}$ is chosen so that the average energy of the
system satisfies
\begin{eqnarray}
  \int^{+\infty}_0 f_{\rm S(D)}(E') (E' - \mu) dE' & = & \frac{(\pi k_{\rm B}
                                                         T_{\rm S(D)})^2}{6},
  \label{a1:eq8}
\end{eqnarray}
which gives $\sigma_{\rm S(D)}=\pi k_{\rm B} T_{\rm S(D)}/\sqrt{3}$. By
combining the Gaussian approximation with Eq.~\eqref{a1:eq1}, the integral
kernel can be approximated via
\begin{eqnarray}
  \Pi_h(\omega) & \approx & \frac{\sqrt{2\pi}}{\sigma_h} e^{-\frac{1}{2} ( \frac{\omega}{\sigma_h} )^2},  
                            \label{a1:eq9}
\end{eqnarray}
with $\sigma_h = \pi k_{\rm B} T_h/\sqrt{3}$ with
$T_h = \sqrt{T^2_{\rm D} - T^2_{\rm S}}$, which is just Eq.~\eqref{s3:eq4}.

\section{Phase control of pulse shaping}
\label{app2}

In this appendix, we introduce the phase control of the pulse shaping following
Ref.~\onlinecite{cook2012}, which we have used in the derivation of
Eq.~\eqref{s3:eq3} from Eq.~\eqref{s3:eq2}.

The problem of phase control of pulse shaping is to find a signal
\begin{eqnarray}
  s(t) & = & a(t) \exp[ i \theta(t) ],
             \label{a2:eq1}
\end{eqnarray}
whose power spectral density follows a given function $U(\omega)$, i.e.,
\begin{eqnarray}
  \hspace{-0.5cm}U(\omega) \exp[ i \Phi(\omega) ] & = & \int^{+\infty}_{-\infty} dt a(t) \exp[
                                                        i \theta(t) - i \omega t],
                                                        \label{a2:eq2}
\end{eqnarray}
with $\Phi(\omega)$ being an arbitrary function. Here $a(t)$, $\theta(t)$,
$\Phi(\omega)$ and $U(\omega)$ are all real.

To attack this problem, one first evaluate the Fourier transformation
Eq.~\eqref{a2:eq2} by using the stationary phase approximation. For a given
$\omega$, this can be done by performing the Taylor expansion of the phase
factor around $t=T_{\omega}$ as
\begin{eqnarray}
   \omega t - \theta(t) & = & [ \omega T_{\omega} - \theta(T_{\omega}) ] + [
                               \omega - \theta'(T_{\omega}) ] (t - T_{\omega})
                              \nonumber\\
                        && \mbox{}+ \theta''(T_{\omega}) (t-T_{\omega})^2 + ...,
                           \label{a2:eq3}
\end{eqnarray}
where $T_{\omega}$ is obtained by using the stationary phase condition
\begin{eqnarray}
  \omega & = & \theta'(T_{\omega}).
               \label{a2:eq4}
\end{eqnarray}
Here $\theta'(t)$ and $\theta''(t)$ represents the first- and second-order
derivative of the function $\theta(t)$ over time $t$, respectively.

Substituting Eqs.~\eqref{a2:eq3} and~\eqref{a2:eq4} into Eq.~\eqref{a2:eq2}, one
obtains 
\begin{eqnarray}
  \hspace{-0.5cm}U(\omega) \exp[ i \Phi(\omega) ] & \approx &
                                                              \sqrt{\frac{2\pi}{\theta''({T_{\omega}})}}
                                                              a(T_{\omega})
  \nonumber\\
  && \mbox{} \times \exp[ -i \omega T_{\omega} + i \theta(T_{\omega}) +
     \frac{\pi}{4} ],
                                                        \label{a2:eq5}
\end{eqnarray}
from which one has
\begin{eqnarray}
  a^2(T_{\omega}) & = & U^2(\omega) \frac{\theta''(T_{\omega}) }{2 \pi}.
                                                        \label{a2:eq6}
\end{eqnarray}

By combining Eq.~\eqref{a2:eq6} with the stationary phase condition
Eq.~\eqref{a2:eq4}, one can obtain a differential equation, from which
$\theta(t)$ can be solved if the function $a(t)$ and $U(\omega)$ are known. To
make this clear, let $x=\omega$ and $y=T_{\omega}$, Eqs.~\eqref{a2:eq4}
and~\eqref{a2:eq6} have the form
\begin{eqnarray}
  x & = & \frac{d\theta(y)}{dy}, \label{a2:eq7}\\
  a^2(y) & = & \frac{U^2(x)}{2\pi} \frac{d^2\theta(y)}{d^2y}.
           \label{a2:eq8}
\end{eqnarray}

Substituting Eq.~\eqref{a2:eq7} into Eq.~\eqref{a2:eq8}, one can eliminate
$\theta(y)$, which gives the differential equation
\begin{eqnarray}
  a^2(y) dy & = & \frac{U^2(x)}{2\pi} dx,
           \label{a2:eq9}
\end{eqnarray}
from which the function $y(x)$[or equivalent, the function $x(y)$] can be
solved. The phase $\theta(y)$ can then be obtained from Eq.~\eqref{a2:eq7}.

To applying the above procedure to solve Eq.~\eqref{s3:eq2}, one choose
\begin{eqnarray}
  a(y) & = & \left\{
             \begin{tabular}{cc}
               $1/\sqrt{t_0}$&, $y \in [0, t_0]$\\
               $0$&, otherwise\\
             \end{tabular}
  \right.,
  \label{a2:eq9}  
\end{eqnarray}
and $U^2(x) = \Pi_h(x)$. Integrating both side of Eq.~\eqref{a2:eq8}, one has
\begin{eqnarray}
  \int^y_{-\infty} a^2(y') dy' & =  & \int^{x}_{-\infty} \frac{d x'}{2\pi} \Pi_h(x).
                      \label{a2:eq10}  
\end{eqnarray}
This equation can be reduced to Eq.~\eqref{s3:eq3} for $y \in [0, t_0]$ by performing the
substitution $y \rightarrow t$ and 
\begin{eqnarray}
  x \rightarrow \frac{e}{\hbar} V(t) = \frac{d\phi(t)}{dt}.
               \label{a2:eq11}
\end{eqnarray}


\begin{thebibliography}{37}%
\makeatletter
\providecommand \@ifxundefined [1]{%
 \@ifx{#1\undefined}
}%
\providecommand \@ifnum [1]{%
 \ifnum #1\expandafter \@firstoftwo
 \else \expandafter \@secondoftwo
 \fi
}%
\providecommand \@ifx [1]{%
 \ifx #1\expandafter \@firstoftwo
 \else \expandafter \@secondoftwo
 \fi
}%
\providecommand \natexlab [1]{#1}%
\providecommand \enquote  [1]{``#1''}%
\providecommand \bibnamefont  [1]{#1}%
\providecommand \bibfnamefont [1]{#1}%
\providecommand \citenamefont [1]{#1}%
\providecommand \href@noop [0]{\@secondoftwo}%
\providecommand \href [0]{\begingroup \@sanitize@url \@href}%
\providecommand \@href[1]{\@@startlink{#1}\@@href}%
\providecommand \@@href[1]{\endgroup#1\@@endlink}%
\providecommand \@sanitize@url [0]{\catcode `\\12\catcode `\$12\catcode
  `\&12\catcode `\#12\catcode `\^12\catcode `\_12\catcode `\%12\relax}%
\providecommand \@@startlink[1]{}%
\providecommand \@@endlink[0]{}%
\providecommand \url  [0]{\begingroup\@sanitize@url \@url }%
\providecommand \@url [1]{\endgroup\@href {#1}{\urlprefix }}%
\providecommand \urlprefix  [0]{URL }%
\providecommand \Eprint [0]{\href }%
\providecommand \doibase [0]{http://dx.doi.org/}%
\providecommand \selectlanguage [0]{\@gobble}%
\providecommand \bibinfo  [0]{\@secondoftwo}%
\providecommand \bibfield  [0]{\@secondoftwo}%
\providecommand \translation [1]{[#1]}%
\providecommand \BibitemOpen [0]{}%
\providecommand \bibitemStop [0]{}%
\providecommand \bibitemNoStop [0]{.\EOS\space}%
\providecommand \EOS [0]{\spacefactor3000\relax}%
\providecommand \BibitemShut  [1]{\csname bibitem#1\endcsname}%
\let\auto@bib@innerbib\@empty
\bibitem [{\citenamefont {Oliver}\ \emph {et~al.}(1999)\citenamefont {Oliver},
  \citenamefont {Kim}, \citenamefont {Liu},\ and\ \citenamefont
  {Yamamoto}}]{oliver1999}%
  \BibitemOpen
  \bibfield  {author} {\bibinfo {author} {\bibfnamefont {W.~D.}\ \bibnamefont
  {Oliver}}, \bibinfo {author} {\bibfnamefont {J.}~\bibnamefont {Kim}},
  \bibinfo {author} {\bibfnamefont {R.~C.}\ \bibnamefont {Liu}}, \ and\
  \bibinfo {author} {\bibfnamefont {Y.}~\bibnamefont {Yamamoto}},\ }\href@noop
  {} {\bibfield  {journal} {\bibinfo  {journal} {Science}\ }\textbf {\bibinfo
  {volume} {284}},\ \bibinfo {pages} {299} (\bibinfo {year}
  {1999})}\BibitemShut {NoStop}%
\bibitem [{\citenamefont {Henny}\ \emph {et~al.}(1999)\citenamefont {Henny},
  \citenamefont {Oberholzer}, \citenamefont {Strunk}, \citenamefont {Heinzel},
  \citenamefont {Ensslin}, \citenamefont {Holland},\ and\ \citenamefont
  {Sch{\"o}nenberger}}]{henny1999}%
  \BibitemOpen
  \bibfield  {author} {\bibinfo {author} {\bibfnamefont {M.}~\bibnamefont
  {Henny}}, \bibinfo {author} {\bibfnamefont {S.}~\bibnamefont {Oberholzer}},
  \bibinfo {author} {\bibfnamefont {C.}~\bibnamefont {Strunk}}, \bibinfo
  {author} {\bibfnamefont {T.}~\bibnamefont {Heinzel}}, \bibinfo {author}
  {\bibfnamefont {K.}~\bibnamefont {Ensslin}}, \bibinfo {author} {\bibfnamefont
  {M.}~\bibnamefont {Holland}}, \ and\ \bibinfo {author} {\bibfnamefont
  {C.}~\bibnamefont {Sch{\"o}nenberger}},\ }\href@noop {} {\bibfield  {journal}
  {\bibinfo  {journal} {Science}\ }\textbf {\bibinfo {volume} {284}},\ \bibinfo
  {pages} {296} (\bibinfo {year} {1999})}\BibitemShut {NoStop}%
\bibitem [{\citenamefont {Bertoni}\ \emph {et~al.}(2000)\citenamefont
  {Bertoni}, \citenamefont {Bordone}, \citenamefont {Brunetti}, \citenamefont
  {Jacoboni},\ and\ \citenamefont {Reggiani}}]{bertoni2000}%
  \BibitemOpen
  \bibfield  {author} {\bibinfo {author} {\bibfnamefont {A.}~\bibnamefont
  {Bertoni}}, \bibinfo {author} {\bibfnamefont {P.}~\bibnamefont {Bordone}},
  \bibinfo {author} {\bibfnamefont {R.}~\bibnamefont {Brunetti}}, \bibinfo
  {author} {\bibfnamefont {C.}~\bibnamefont {Jacoboni}}, \ and\ \bibinfo
  {author} {\bibfnamefont {S.}~\bibnamefont {Reggiani}},\ }\href@noop {}
  {\bibfield  {journal} {\bibinfo  {journal} {Phys. Rev. Lett.}\ }\textbf
  {\bibinfo {volume} {84}},\ \bibinfo {pages} {5912} (\bibinfo {year}
  {2000})}\BibitemShut {NoStop}%
\bibitem [{\citenamefont {Ionicioiu}\ \emph {et~al.}(2001)\citenamefont
  {Ionicioiu}, \citenamefont {Amaratunga},\ and\ \citenamefont
  {Udrea}}]{ionicioiu2001}%
  \BibitemOpen
  \bibfield  {author} {\bibinfo {author} {\bibfnamefont {R.}~\bibnamefont
  {Ionicioiu}}, \bibinfo {author} {\bibfnamefont {G.}~\bibnamefont
  {Amaratunga}}, \ and\ \bibinfo {author} {\bibfnamefont {F.}~\bibnamefont
  {Udrea}},\ }\href@noop {} {\bibfield  {journal} {\bibinfo  {journal} {Int. J.
  Mod. Phys. B}\ }\textbf {\bibinfo {volume} {15}},\ \bibinfo {pages} {125}
  (\bibinfo {year} {2001})}\BibitemShut {NoStop}%
\bibitem [{\citenamefont {Ji}\ \emph {et~al.}(2003)\citenamefont {Ji},
  \citenamefont {Chung}, \citenamefont {Sprinzak}, \citenamefont {Heiblum},
  \citenamefont {Mahalu},\ and\ \citenamefont {Shtrikman}}]{ji2003}%
  \BibitemOpen
  \bibfield  {author} {\bibinfo {author} {\bibfnamefont {Y.}~\bibnamefont
  {Ji}}, \bibinfo {author} {\bibfnamefont {Y.}~\bibnamefont {Chung}}, \bibinfo
  {author} {\bibfnamefont {D.}~\bibnamefont {Sprinzak}}, \bibinfo {author}
  {\bibfnamefont {M.}~\bibnamefont {Heiblum}}, \bibinfo {author} {\bibfnamefont
  {D.}~\bibnamefont {Mahalu}}, \ and\ \bibinfo {author} {\bibfnamefont
  {H.}~\bibnamefont {Shtrikman}},\ }\href@noop {} {\bibfield  {journal}
  {\bibinfo  {journal} {Nature}\ }\textbf {\bibinfo {volume} {422}},\ \bibinfo
  {pages} {415} (\bibinfo {year} {2003})}\BibitemShut {NoStop}%
\bibitem [{\citenamefont {Bocquillon}\ \emph {et~al.}(2014)\citenamefont
  {Bocquillon}, \citenamefont {Freulon}, \citenamefont {Parmentier},
  \citenamefont {Berroir}, \citenamefont {Pla{\c{c}}ais}, \citenamefont {Wahl},
  \citenamefont {Rech}, \citenamefont {Jonckheere}, \citenamefont {Martin},
  \citenamefont {Grenier}, \citenamefont {Ferraro}, \citenamefont
  {Degiovanni},\ and\ \citenamefont {F{\`e}ve}}]{bocquillon2014}%
  \BibitemOpen
  \bibfield  {author} {\bibinfo {author} {\bibfnamefont {E.}~\bibnamefont
  {Bocquillon}}, \bibinfo {author} {\bibfnamefont {V.}~\bibnamefont {Freulon}},
  \bibinfo {author} {\bibfnamefont {F.~D.}\ \bibnamefont {Parmentier}},
  \bibinfo {author} {\bibfnamefont {J.-M.}\ \bibnamefont {Berroir}}, \bibinfo
  {author} {\bibfnamefont {B.}~\bibnamefont {Pla{\c{c}}ais}}, \bibinfo {author}
  {\bibfnamefont {C.}~\bibnamefont {Wahl}}, \bibinfo {author} {\bibfnamefont
  {J.}~\bibnamefont {Rech}}, \bibinfo {author} {\bibfnamefont {T.}~\bibnamefont
  {Jonckheere}}, \bibinfo {author} {\bibfnamefont {T.}~\bibnamefont {Martin}},
  \bibinfo {author} {\bibfnamefont {C.}~\bibnamefont {Grenier}}, \bibinfo
  {author} {\bibfnamefont {D.}~\bibnamefont {Ferraro}}, \bibinfo {author}
  {\bibfnamefont {P.}~\bibnamefont {Degiovanni}}, \ and\ \bibinfo {author}
  {\bibfnamefont {G.}~\bibnamefont {F{\`e}ve}},\ }\href@noop {} {\bibfield
  {journal} {\bibinfo  {journal} {Ann. Phys.}\ }\textbf {\bibinfo {volume}
  {526}},\ \bibinfo {pages} {1} (\bibinfo {year} {2014})}\BibitemShut {NoStop}%
\bibitem [{\citenamefont {Glattli}\ and\ \citenamefont
  {Roulleau}(2017)}]{glattli2017}%
  \BibitemOpen
  \bibfield  {author} {\bibinfo {author} {\bibfnamefont {D.~C.}\ \bibnamefont
  {Glattli}}\ and\ \bibinfo {author} {\bibfnamefont {P.~S.}\ \bibnamefont
  {Roulleau}},\ }\href@noop {} {\bibfield  {journal} {\bibinfo  {journal}
  {Phys. Stat. Sol. (b)}\ }\textbf {\bibinfo {volume} {254}} (\bibinfo {year}
  {2017})}\BibitemShut {NoStop}%
\bibitem [{\citenamefont {Vanevi{\'c}}\ \emph {et~al.}(2016)\citenamefont
  {Vanevi{\'c}}, \citenamefont {Gabelli}, \citenamefont {Belzig},\ and\
  \citenamefont {Reulet}}]{vanevic2016}%
  \BibitemOpen
  \bibfield  {author} {\bibinfo {author} {\bibfnamefont {M.}~\bibnamefont
  {Vanevi{\'c}}}, \bibinfo {author} {\bibfnamefont {J.}~\bibnamefont
  {Gabelli}}, \bibinfo {author} {\bibfnamefont {W.}~\bibnamefont {Belzig}}, \
  and\ \bibinfo {author} {\bibfnamefont {B.}~\bibnamefont {Reulet}},\
  }\href@noop {} {\bibfield  {journal} {\bibinfo  {journal} {Phys. Rev. B}\
  }\textbf {\bibinfo {volume} {93}},\ \bibinfo {pages} {041416} (\bibinfo
  {year} {2016})}\BibitemShut {NoStop}%
\bibitem [{\citenamefont {Lee}\ and\ \citenamefont {Levitov}(1993)}]{lee1993}%
  \BibitemOpen
  \bibfield  {author} {\bibinfo {author} {\bibfnamefont {H.}~\bibnamefont
  {Lee}}\ and\ \bibinfo {author} {\bibfnamefont {L.}~\bibnamefont {Levitov}},\
  }\href@noop {} {\bibfield  {journal} {\bibinfo  {journal} {cond-mat/9312013}\
  } (\bibinfo {year} {1993})}\BibitemShut {NoStop}%
\bibitem [{\citenamefont {Lee}\ and\ \citenamefont {Levitov}(1995)}]{lee1995}%
  \BibitemOpen
  \bibfield  {author} {\bibinfo {author} {\bibfnamefont {H.}~\bibnamefont
  {Lee}}\ and\ \bibinfo {author} {\bibfnamefont {L.}~\bibnamefont {Levitov}},\
  }\href@noop {} {\bibfield  {journal} {\bibinfo  {journal} {cond-mat/9507011}\
  } (\bibinfo {year} {1995})}\BibitemShut {NoStop}%
\bibitem [{\citenamefont {Keeling}\ \emph {et~al.}(2006)\citenamefont
  {Keeling}, \citenamefont {Klich},\ and\ \citenamefont
  {Levitov}}]{keeling2006}%
  \BibitemOpen
  \bibfield  {author} {\bibinfo {author} {\bibfnamefont {J.}~\bibnamefont
  {Keeling}}, \bibinfo {author} {\bibfnamefont {I.}~\bibnamefont {Klich}}, \
  and\ \bibinfo {author} {\bibfnamefont {L.}~\bibnamefont {Levitov}},\
  }\href@noop {} {\bibfield  {journal} {\bibinfo  {journal} {Phys. Rev. Lett.}\
  }\textbf {\bibinfo {volume} {97}},\ \bibinfo {pages} {116403} (\bibinfo
  {year} {2006})}\BibitemShut {NoStop}%
\bibitem [{\citenamefont {Dubois}\ \emph {et~al.}(2013)\citenamefont {Dubois},
  \citenamefont {Jullien}, \citenamefont {Portier}, \citenamefont {Roche},
  \citenamefont {Cavanna}, \citenamefont {Jin}, \citenamefont {Wegscheider},
  \citenamefont {Roulleau},\ and\ \citenamefont {Glattli}}]{dubois2013}%
  \BibitemOpen
  \bibfield  {author} {\bibinfo {author} {\bibfnamefont {J.}~\bibnamefont
  {Dubois}}, \bibinfo {author} {\bibfnamefont {T.}~\bibnamefont {Jullien}},
  \bibinfo {author} {\bibfnamefont {F.}~\bibnamefont {Portier}}, \bibinfo
  {author} {\bibfnamefont {P.}~\bibnamefont {Roche}}, \bibinfo {author}
  {\bibfnamefont {A.}~\bibnamefont {Cavanna}}, \bibinfo {author} {\bibfnamefont
  {Y.}~\bibnamefont {Jin}}, \bibinfo {author} {\bibfnamefont {W.}~\bibnamefont
  {Wegscheider}}, \bibinfo {author} {\bibfnamefont {P.}~\bibnamefont
  {Roulleau}}, \ and\ \bibinfo {author} {\bibfnamefont {D.}~\bibnamefont
  {Glattli}},\ }\href@noop {} {\bibfield  {journal} {\bibinfo  {journal}
  {Nature}\ }\textbf {\bibinfo {volume} {502}},\ \bibinfo {pages} {659}
  (\bibinfo {year} {2013})}\BibitemShut {NoStop}%
\bibitem [{\citenamefont {Jullien}\ \emph {et~al.}(2014)\citenamefont
  {Jullien}, \citenamefont {Roulleau}, \citenamefont {Roche}, \citenamefont
  {Cavanna}, \citenamefont {Jin},\ and\ \citenamefont {Glattli}}]{jullien2014}%
  \BibitemOpen
  \bibfield  {author} {\bibinfo {author} {\bibfnamefont {T.}~\bibnamefont
  {Jullien}}, \bibinfo {author} {\bibfnamefont {P.}~\bibnamefont {Roulleau}},
  \bibinfo {author} {\bibfnamefont {B.}~\bibnamefont {Roche}}, \bibinfo
  {author} {\bibfnamefont {A.}~\bibnamefont {Cavanna}}, \bibinfo {author}
  {\bibfnamefont {Y.}~\bibnamefont {Jin}}, \ and\ \bibinfo {author}
  {\bibfnamefont {D.}~\bibnamefont {Glattli}},\ }\href@noop {} {\bibfield
  {journal} {\bibinfo  {journal} {Nature}\ }\textbf {\bibinfo {volume} {514}},\
  \bibinfo {pages} {603} (\bibinfo {year} {2014})}\BibitemShut {NoStop}%
\bibitem [{\citenamefont {Gabelli}\ \emph {et~al.}(2017)\citenamefont
  {Gabelli}, \citenamefont {Thibault}, \citenamefont {Gasse}, \citenamefont
  {Lupien},\ and\ \citenamefont {Reulet}}]{gabelli2017}%
  \BibitemOpen
  \bibfield  {author} {\bibinfo {author} {\bibfnamefont {J.}~\bibnamefont
  {Gabelli}}, \bibinfo {author} {\bibfnamefont {K.}~\bibnamefont {Thibault}},
  \bibinfo {author} {\bibfnamefont {G.}~\bibnamefont {Gasse}}, \bibinfo
  {author} {\bibfnamefont {C.}~\bibnamefont {Lupien}}, \ and\ \bibinfo {author}
  {\bibfnamefont {B.}~\bibnamefont {Reulet}},\ }\href@noop {} {\bibfield
  {journal} {\bibinfo  {journal} {Phys. Stat. Sol. (b)}\ }\textbf {\bibinfo
  {volume} {254}} (\bibinfo {year} {2017})}\BibitemShut {NoStop}%
\bibitem [{\citenamefont {Moskalets}(2016)}]{moskalets2016}%
  \BibitemOpen
  \bibfield  {author} {\bibinfo {author} {\bibfnamefont {M.}~\bibnamefont
  {Moskalets}},\ }\href@noop {} {\bibfield  {journal} {\bibinfo  {journal}
  {Phys. Rev. Lett.}\ }\textbf {\bibinfo {volume} {117}},\ \bibinfo {pages}
  {046801} (\bibinfo {year} {2016})}\BibitemShut {NoStop}%
\bibitem [{\citenamefont {Schwab}\ \emph {et~al.}(2000)\citenamefont {Schwab},
  \citenamefont {Henriksen}, \citenamefont {Worlock},\ and\ \citenamefont
  {Roukes}}]{schwab2000}%
  \BibitemOpen
  \bibfield  {author} {\bibinfo {author} {\bibfnamefont {K.}~\bibnamefont
  {Schwab}}, \bibinfo {author} {\bibfnamefont {E.~A.}\ \bibnamefont
  {Henriksen}}, \bibinfo {author} {\bibfnamefont {J.~M.}\ \bibnamefont
  {Worlock}}, \ and\ \bibinfo {author} {\bibfnamefont {M.~L.}\ \bibnamefont
  {Roukes}},\ }\href
  {https://search.proquest.com/docview/204486250?accountid=13854} {\bibfield
  {journal} {\bibinfo  {journal} {Nature}\ }\textbf {\bibinfo {volume} {404}},\
  \bibinfo {pages} {974} (\bibinfo {year} {2000})}\BibitemShut {NoStop}%
\bibitem [{\citenamefont {Chiatti}\ \emph {et~al.}(2006)\citenamefont
  {Chiatti}, \citenamefont {Nicholls}, \citenamefont {Proskuryakov},
  \citenamefont {Lumpkin}, \citenamefont {Farrer},\ and\ \citenamefont
  {Ritchie}}]{chiatti2006}%
  \BibitemOpen
  \bibfield  {author} {\bibinfo {author} {\bibfnamefont {O.}~\bibnamefont
  {Chiatti}}, \bibinfo {author} {\bibfnamefont {J.~T.}\ \bibnamefont
  {Nicholls}}, \bibinfo {author} {\bibfnamefont {Y.~Y.}\ \bibnamefont
  {Proskuryakov}}, \bibinfo {author} {\bibfnamefont {N.}~\bibnamefont
  {Lumpkin}}, \bibinfo {author} {\bibfnamefont {I.}~\bibnamefont {Farrer}}, \
  and\ \bibinfo {author} {\bibfnamefont {D.~A.}\ \bibnamefont {Ritchie}},\
  }\href {\doibase 10.1103/PhysRevLett.97.056601} {\bibfield  {journal}
  {\bibinfo  {journal} {Phys. Rev. Lett.}\ }\textbf {\bibinfo {volume} {97}},\
  \bibinfo {pages} {056601} (\bibinfo {year} {2006})}\BibitemShut {NoStop}%
\bibitem [{\citenamefont {Chen}\ \emph {et~al.}(2009)\citenamefont {Chen},
  \citenamefont {Dirks}, \citenamefont {Al-Zoubi}, \citenamefont {Birge},\ and\
  \citenamefont {Mason}}]{chen2009}%
  \BibitemOpen
  \bibfield  {author} {\bibinfo {author} {\bibfnamefont {Y.-F.}\ \bibnamefont
  {Chen}}, \bibinfo {author} {\bibfnamefont {T.}~\bibnamefont {Dirks}},
  \bibinfo {author} {\bibfnamefont {G.}~\bibnamefont {Al-Zoubi}}, \bibinfo
  {author} {\bibfnamefont {N.~O.}\ \bibnamefont {Birge}}, \ and\ \bibinfo
  {author} {\bibfnamefont {N.}~\bibnamefont {Mason}},\ }\href@noop {}
  {\bibfield  {journal} {\bibinfo  {journal} {Phys. Rev. Lett.}\ }\textbf
  {\bibinfo {volume} {102}},\ \bibinfo {pages} {036804} (\bibinfo {year}
  {2009})}\BibitemShut {NoStop}%
\bibitem [{\citenamefont {Granger}\ \emph {et~al.}(2009)\citenamefont
  {Granger}, \citenamefont {Eisenstein},\ and\ \citenamefont
  {Reno}}]{granger2009}%
  \BibitemOpen
  \bibfield  {author} {\bibinfo {author} {\bibfnamefont {G.}~\bibnamefont
  {Granger}}, \bibinfo {author} {\bibfnamefont {J.}~\bibnamefont {Eisenstein}},
  \ and\ \bibinfo {author} {\bibfnamefont {J.}~\bibnamefont {Reno}},\
  }\href@noop {} {\bibfield  {journal} {\bibinfo  {journal} {Phys. Rev. Lett.}\
  }\textbf {\bibinfo {volume} {102}},\ \bibinfo {pages} {086803} (\bibinfo
  {year} {2009})}\BibitemShut {NoStop}%
\bibitem [{\citenamefont {Altimiras}\ \emph {et~al.}(2010)\citenamefont
  {Altimiras}, \citenamefont {Le~Sueur}, \citenamefont {Gennser}, \citenamefont
  {Cavanna}, \citenamefont {Mailly},\ and\ \citenamefont
  {Pierre}}]{altimiras2010}%
  \BibitemOpen
  \bibfield  {author} {\bibinfo {author} {\bibfnamefont {C.}~\bibnamefont
  {Altimiras}}, \bibinfo {author} {\bibfnamefont {H.}~\bibnamefont {Le~Sueur}},
  \bibinfo {author} {\bibfnamefont {U.}~\bibnamefont {Gennser}}, \bibinfo
  {author} {\bibfnamefont {A.}~\bibnamefont {Cavanna}}, \bibinfo {author}
  {\bibfnamefont {D.}~\bibnamefont {Mailly}}, \ and\ \bibinfo {author}
  {\bibfnamefont {F.}~\bibnamefont {Pierre}},\ }\href@noop {} {\bibfield
  {journal} {\bibinfo  {journal} {Nature Physics}\ }\textbf {\bibinfo {volume}
  {6}},\ \bibinfo {pages} {34} (\bibinfo {year} {2010})}\BibitemShut {NoStop}%
\bibitem [{\citenamefont {Le~Sueur}\ \emph {et~al.}(2010)\citenamefont
  {Le~Sueur}, \citenamefont {Altimiras}, \citenamefont {Gennser}, \citenamefont
  {Cavanna}, \citenamefont {Mailly},\ and\ \citenamefont {Pierre}}]{le2010}%
  \BibitemOpen
  \bibfield  {author} {\bibinfo {author} {\bibfnamefont {H.}~\bibnamefont
  {Le~Sueur}}, \bibinfo {author} {\bibfnamefont {C.}~\bibnamefont {Altimiras}},
  \bibinfo {author} {\bibfnamefont {U.}~\bibnamefont {Gennser}}, \bibinfo
  {author} {\bibfnamefont {A.}~\bibnamefont {Cavanna}}, \bibinfo {author}
  {\bibfnamefont {D.}~\bibnamefont {Mailly}}, \ and\ \bibinfo {author}
  {\bibfnamefont {F.}~\bibnamefont {Pierre}},\ }\href@noop {} {\bibfield
  {journal} {\bibinfo  {journal} {Phys. Rev. Lett.}\ }\textbf {\bibinfo
  {volume} {105}},\ \bibinfo {pages} {056803} (\bibinfo {year}
  {2010})}\BibitemShut {NoStop}%
\bibitem [{\citenamefont {Venkatachalam}\ \emph {et~al.}(2012)\citenamefont
  {Venkatachalam}, \citenamefont {Hart}, \citenamefont {Pfeiffer},
  \citenamefont {West},\ and\ \citenamefont {Yacoby}}]{venkatachalam2012}%
  \BibitemOpen
  \bibfield  {author} {\bibinfo {author} {\bibfnamefont {V.}~\bibnamefont
  {Venkatachalam}}, \bibinfo {author} {\bibfnamefont {S.}~\bibnamefont {Hart}},
  \bibinfo {author} {\bibfnamefont {L.}~\bibnamefont {Pfeiffer}}, \bibinfo
  {author} {\bibfnamefont {K.}~\bibnamefont {West}}, \ and\ \bibinfo {author}
  {\bibfnamefont {A.}~\bibnamefont {Yacoby}},\ }\href@noop {} {\bibfield
  {journal} {\bibinfo  {journal} {Nature Physics}\ }\textbf {\bibinfo {volume}
  {8}} (\bibinfo {year} {2012})}\BibitemShut {NoStop}%
\bibitem [{\citenamefont {Jezouin}\ \emph {et~al.}(2013)\citenamefont
  {Jezouin}, \citenamefont {Parmentier}, \citenamefont {Anthore}, \citenamefont
  {Gennser}, \citenamefont {Cavanna}, \citenamefont {Jin},\ and\ \citenamefont
  {Pierre}}]{Jezouin2013}%
  \BibitemOpen
  \bibfield  {author} {\bibinfo {author} {\bibfnamefont {S.}~\bibnamefont
  {Jezouin}}, \bibinfo {author} {\bibfnamefont {F.~D.}\ \bibnamefont
  {Parmentier}}, \bibinfo {author} {\bibfnamefont {A.}~\bibnamefont {Anthore}},
  \bibinfo {author} {\bibfnamefont {U.}~\bibnamefont {Gennser}}, \bibinfo
  {author} {\bibfnamefont {A.}~\bibnamefont {Cavanna}}, \bibinfo {author}
  {\bibfnamefont {Y.}~\bibnamefont {Jin}}, \ and\ \bibinfo {author}
  {\bibfnamefont {F.}~\bibnamefont {Pierre}},\ }\href {\doibase
  10.1126/science.1241912} {\bibfield  {journal} {\bibinfo  {journal}
  {Science}\ }\textbf {\bibinfo {volume} {342}},\ \bibinfo {pages} {601}
  (\bibinfo {year} {2013})}\BibitemShut {NoStop}%
\bibitem [{\citenamefont {Spietz}\ \emph {et~al.}(2003)\citenamefont {Spietz},
  \citenamefont {Lehnert}, \citenamefont {Siddiqi},\ and\ \citenamefont
  {Schoelkopf}}]{spietz2003}%
  \BibitemOpen
  \bibfield  {author} {\bibinfo {author} {\bibfnamefont {L.}~\bibnamefont
  {Spietz}}, \bibinfo {author} {\bibfnamefont {K.}~\bibnamefont {Lehnert}},
  \bibinfo {author} {\bibfnamefont {I.}~\bibnamefont {Siddiqi}}, \ and\
  \bibinfo {author} {\bibfnamefont {R.}~\bibnamefont {Schoelkopf}},\
  }\href@noop {} {\bibfield  {journal} {\bibinfo  {journal} {Science}\ }\textbf
  {\bibinfo {volume} {300}},\ \bibinfo {pages} {1929} (\bibinfo {year}
  {2003})}\BibitemShut {NoStop}%
\bibitem [{\citenamefont {Blanter}\ and\ \citenamefont
  {B{\"u}ttiker}(2000)}]{blanter2000}%
  \BibitemOpen
  \bibfield  {author} {\bibinfo {author} {\bibfnamefont {Y.~M.}\ \bibnamefont
  {Blanter}}\ and\ \bibinfo {author} {\bibfnamefont {M.}~\bibnamefont
  {B{\"u}ttiker}},\ }\href@noop {} {\bibfield  {journal} {\bibinfo  {journal}
  {Physics Reports}\ }\textbf {\bibinfo {volume} {336}},\ \bibinfo {pages} {1}
  (\bibinfo {year} {2000})}\BibitemShut {NoStop}%
\bibitem [{\citenamefont {Ivanov}\ \emph {et~al.}(1997)\citenamefont {Ivanov},
  \citenamefont {Lee},\ and\ \citenamefont {Levitov}}]{ivanov1997}%
  \BibitemOpen
  \bibfield  {author} {\bibinfo {author} {\bibfnamefont {D.}~\bibnamefont
  {Ivanov}}, \bibinfo {author} {\bibfnamefont {H.}~\bibnamefont {Lee}}, \ and\
  \bibinfo {author} {\bibfnamefont {L.}~\bibnamefont {Levitov}},\ }\href@noop
  {} {\bibfield  {journal} {\bibinfo  {journal} {Phys. Rev. B}\ }\textbf
  {\bibinfo {volume} {56}},\ \bibinfo {pages} {6839} (\bibinfo {year}
  {1997})}\BibitemShut {NoStop}%
\bibitem [{\citenamefont {Keeling}\ \emph {et~al.}(2008)\citenamefont
  {Keeling}, \citenamefont {Shytov},\ and\ \citenamefont
  {Levitov}}]{keeling2008}%
  \BibitemOpen
  \bibfield  {author} {\bibinfo {author} {\bibfnamefont {J.}~\bibnamefont
  {Keeling}}, \bibinfo {author} {\bibfnamefont {A.}~\bibnamefont {Shytov}}, \
  and\ \bibinfo {author} {\bibfnamefont {L.}~\bibnamefont {Levitov}},\
  }\href@noop {} {\bibfield  {journal} {\bibinfo  {journal} {Phys. Rev. Lett.}\
  }\textbf {\bibinfo {volume} {101}},\ \bibinfo {pages} {196404} (\bibinfo
  {year} {2008})}\BibitemShut {NoStop}%
\bibitem [{\citenamefont {Grenier}\ \emph {et~al.}(2011)\citenamefont
  {Grenier}, \citenamefont {Herv{\'e}}, \citenamefont {Bocquillon},
  \citenamefont {Parmentier}, \citenamefont {Pla{\c{c}}ais}, \citenamefont
  {Berroir}, \citenamefont {F{\`e}ve},\ and\ \citenamefont
  {Degiovanni}}]{grenier2011}%
  \BibitemOpen
  \bibfield  {author} {\bibinfo {author} {\bibfnamefont {C.}~\bibnamefont
  {Grenier}}, \bibinfo {author} {\bibfnamefont {R.}~\bibnamefont {Herv{\'e}}},
  \bibinfo {author} {\bibfnamefont {E.}~\bibnamefont {Bocquillon}}, \bibinfo
  {author} {\bibfnamefont {F.~D.}\ \bibnamefont {Parmentier}}, \bibinfo
  {author} {\bibfnamefont {B.}~\bibnamefont {Pla{\c{c}}ais}}, \bibinfo {author}
  {\bibfnamefont {J.-M.}\ \bibnamefont {Berroir}}, \bibinfo {author}
  {\bibfnamefont {G.}~\bibnamefont {F{\`e}ve}}, \ and\ \bibinfo {author}
  {\bibfnamefont {P.}~\bibnamefont {Degiovanni}},\ }\href@noop {} {\bibfield
  {journal} {\bibinfo  {journal} {New Journal of Physics}\ }\textbf {\bibinfo
  {volume} {13}},\ \bibinfo {pages} {093007} (\bibinfo {year}
  {2011})}\BibitemShut {NoStop}%
\bibitem [{\citenamefont {Haack}\ \emph {et~al.}(2011)\citenamefont {Haack},
  \citenamefont {Moskalets}, \citenamefont {Splettstoesser},\ and\
  \citenamefont {B{\"u}ttiker}}]{haack2011}%
  \BibitemOpen
  \bibfield  {author} {\bibinfo {author} {\bibfnamefont {G.}~\bibnamefont
  {Haack}}, \bibinfo {author} {\bibfnamefont {M.}~\bibnamefont {Moskalets}},
  \bibinfo {author} {\bibfnamefont {J.}~\bibnamefont {Splettstoesser}}, \ and\
  \bibinfo {author} {\bibfnamefont {M.}~\bibnamefont {B{\"u}ttiker}},\
  }\href@noop {} {\bibfield  {journal} {\bibinfo  {journal} {Phys. Rev. B}\
  }\textbf {\bibinfo {volume} {84}},\ \bibinfo {pages} {081303} (\bibinfo
  {year} {2011})}\BibitemShut {NoStop}%
\bibitem [{\citenamefont {Haack}\ \emph {et~al.}(2013)\citenamefont {Haack},
  \citenamefont {Moskalets},\ and\ \citenamefont {B{\"u}ttiker}}]{haack2013}%
  \BibitemOpen
  \bibfield  {author} {\bibinfo {author} {\bibfnamefont {G.}~\bibnamefont
  {Haack}}, \bibinfo {author} {\bibfnamefont {M.}~\bibnamefont {Moskalets}}, \
  and\ \bibinfo {author} {\bibfnamefont {M.}~\bibnamefont {B{\"u}ttiker}},\
  }\href@noop {} {\bibfield  {journal} {\bibinfo  {journal} {Phys. Rev. B}\
  }\textbf {\bibinfo {volume} {87}},\ \bibinfo {pages} {201302} (\bibinfo
  {year} {2013})}\BibitemShut {NoStop}%
\bibitem [{\citenamefont {Kovrizhin}\ and\ \citenamefont
  {Chalker}(2011)}]{kovrizhin2011}%
  \BibitemOpen
  \bibfield  {author} {\bibinfo {author} {\bibfnamefont {D.}~\bibnamefont
  {Kovrizhin}}\ and\ \bibinfo {author} {\bibfnamefont {J.}~\bibnamefont
  {Chalker}},\ }\href@noop {} {\bibfield  {journal} {\bibinfo  {journal} {Phys.
  Rev. B}\ }\textbf {\bibinfo {volume} {84}},\ \bibinfo {pages} {085105}
  (\bibinfo {year} {2011})}\BibitemShut {NoStop}%
\bibitem [{\citenamefont {Kovrizhin}\ and\ \citenamefont
  {Chalker}(2012)}]{kovrizhin2012}%
  \BibitemOpen
  \bibfield  {author} {\bibinfo {author} {\bibfnamefont {D.}~\bibnamefont
  {Kovrizhin}}\ and\ \bibinfo {author} {\bibfnamefont {J.}~\bibnamefont
  {Chalker}},\ }\href@noop {} {\bibfield  {journal} {\bibinfo  {journal} {Phys.
  Rev. Lett.}\ }\textbf {\bibinfo {volume} {109}},\ \bibinfo {pages} {106403}
  (\bibinfo {year} {2012})}\BibitemShut {NoStop}%
\bibitem [{\citenamefont {Moskalets}(2014)}]{moskalets2014}%
  \BibitemOpen
  \bibfield  {author} {\bibinfo {author} {\bibfnamefont {M.}~\bibnamefont
  {Moskalets}},\ }\href@noop {} {\bibfield  {journal} {\bibinfo  {journal}
  {Phys. Rev. B}\ }\textbf {\bibinfo {volume} {89}},\ \bibinfo {pages} {045402}
  (\bibinfo {year} {2014})}\BibitemShut {NoStop}%
\bibitem [{\citenamefont {Weiner}(2011)}]{weiner2011}%
  \BibitemOpen
  \bibfield  {author} {\bibinfo {author} {\bibfnamefont {A.~M.}\ \bibnamefont
  {Weiner}},\ }\href@noop {} {\bibfield  {journal} {\bibinfo  {journal} {Optics
  Communications}\ }\textbf {\bibinfo {volume} {284}},\ \bibinfo {pages} {3669}
  (\bibinfo {year} {2011})}\BibitemShut {NoStop}%
\bibitem [{\citenamefont {Kumar}\ \emph {et~al.}(1996)\citenamefont {Kumar},
  \citenamefont {Saminadayar}, \citenamefont {Glattli}, \citenamefont {Jin},\
  and\ \citenamefont {Etienne}}]{kumar1996}%
  \BibitemOpen
  \bibfield  {author} {\bibinfo {author} {\bibfnamefont {A.}~\bibnamefont
  {Kumar}}, \bibinfo {author} {\bibfnamefont {L.}~\bibnamefont {Saminadayar}},
  \bibinfo {author} {\bibfnamefont {D.}~\bibnamefont {Glattli}}, \bibinfo
  {author} {\bibfnamefont {Y.}~\bibnamefont {Jin}}, \ and\ \bibinfo {author}
  {\bibfnamefont {B.}~\bibnamefont {Etienne}},\ }\href@noop {} {\bibfield
  {journal} {\bibinfo  {journal} {Phys. Rev. Lett.}\ }\textbf {\bibinfo
  {volume} {76}},\ \bibinfo {pages} {2778} (\bibinfo {year}
  {1996})}\BibitemShut {NoStop}%
\bibitem [{\citenamefont {Prokudina}\ \emph {et~al.}(2014)\citenamefont
  {Prokudina}, \citenamefont {Ludwig}, \citenamefont {Pellegrini},
  \citenamefont {Sorba}, \citenamefont {Biasiol},\ and\ \citenamefont
  {Khrapai}}]{prokudina2014}%
  \BibitemOpen
  \bibfield  {author} {\bibinfo {author} {\bibfnamefont {M.}~\bibnamefont
  {Prokudina}}, \bibinfo {author} {\bibfnamefont {S.}~\bibnamefont {Ludwig}},
  \bibinfo {author} {\bibfnamefont {V.}~\bibnamefont {Pellegrini}}, \bibinfo
  {author} {\bibfnamefont {L.}~\bibnamefont {Sorba}}, \bibinfo {author}
  {\bibfnamefont {G.}~\bibnamefont {Biasiol}}, \ and\ \bibinfo {author}
  {\bibfnamefont {V.}~\bibnamefont {Khrapai}},\ }\href@noop {} {\bibfield
  {journal} {\bibinfo  {journal} {Phys. Rev. Lett.}\ }\textbf {\bibinfo
  {volume} {112}},\ \bibinfo {pages} {216402} (\bibinfo {year}
  {2014})}\BibitemShut {NoStop}%
\bibitem [{\citenamefont {Cook}(2012)}]{cook2012}%
  \BibitemOpen
  \bibfield  {author} {\bibinfo {author} {\bibfnamefont {C.}~\bibnamefont
  {Cook}},\ }\href@noop {} {\emph {\bibinfo {title} {Radar signals: An
  introduction to theory and application}}}\ (\bibinfo  {publisher}
  {Elsevier},\ \bibinfo {year} {2012})\BibitemShut {NoStop}%
\end{thebibliography}

%

\end{document}